\documentclass[9pt,twocolumn,twoside]{osajnl}
\journal{josab}

\usepackage{bbm}
\usepackage{braket}

\DeclareMathOperator{\trace}{tr} 

\title{The role of diffraction in the Casimir effect beyond the proximity force approximation}

\author[1,*]{Vinicius Henning}
\author[1,2]{Benjamin Spreng}
\author[2]{Michael Hartmann}
\author[2]{Gert-Ludwig Ingold}
\author[1]{Paulo A. Maia Neto}

\affil[1]{Instituto de F{\'i}sica, Universidade Federal do Rio de Janeiro, CP 68528, Rio de Janeiro RJ 21941-909, Brazil}
\affil[2]{Universität Augsburg, Institut für Physik, 86135 Augsburg, Germany}

\affil[*]{Corresponding author: henning@if.ufrj.br}

\date{\today}

\begin{abstract}
We derive the leading-order correction to the proximity force approximation
(PFA) result for the electromagnetic Casimir interaction in the plane-sphere
geometry by developing the scattering approach in the plane-wave basis.
Expressing the Casimir energy as a sum over round trips between plane and
sphere, we find two distinct contributions to the correction. The first one
results from the variation of the Mie reflection operator, calculated within the
geometric optical WKB approximation, over the narrow Fourier interval associated to
specular reflection at the vicinity of the point of closest approach on the
spherical surface. The second contribution, accounting for roughly 90\% of
the total correction, results from the modification of the geometric optical WKB Mie
scattering amplitude due to diffraction. Our derivation provides a clear physical
understanding of the nature of the PFA correction for spherical surfaces.
\end{abstract}

\setboolean{displaycopyright}{false}

\begin{document}

\maketitle

\section{Introduction}
\label{sec:introduction}

The Casimir interaction between material surfaces is a striking consequence of
the quantum nature of electrodynamics \cite{Casimir1948}. The plane-sphere
geometry is particularly suited to implement very precise measurements of the
Casimir force or force gradient
\cite{Bordag2009,Klimchitskaya2009,Decca2011,Lamoreaux2011}. The interaction
for such geometry has been probed for a number of different materials over the
last decade
\cite{Sushkov2011,Torricelli2011,Chang2012,GarciaSanchez2012,Banishev2013,Sedighi2016,Bimonte2016}.
The Casimir interaction between two spherical surfaces is also of great
interest \cite{Elzbieciak-Wodka2014,Ether2015,Garrett2018} given its
applications in colloids and surface sciences \cite{Butt2010}.

Until recently, the theoretical description of plane-sphere experiments has
been limited to the employment of the proximity force approximation (PFA), also
known as Derjaguin approximation \cite{Derjaguin1934}. Within PFA, the Casimir
energy is obtained from Lifshitz's formula for parallel planes by averaging
over the local distance between the surfaces \cite{Parsegian2006}. PFA provides
the correct leading asymptotics for large sphere radius $R$ when considering
general materials and arbitrary temperatures \cite{Spreng2018}.  As reviewed in
\cite{Hartmann2018}, typical experiments are close to the validity range of PFA
since they correspond to aspect ratios $R/L> 10^2,$ where $L$ is the minimal
distance between the surfaces as indicated in Fig.~1.  Nevertheless, it is
still necessary to access the accuracy of this approximation when comparing
with experimental data. Early
attempts~\cite{MaiaNeto2008,Emig2008,Canaguier-Durand2009,Canaguier-Durand2010,Canaguier2010PRA,Zandi2010}
to derive exact numerical results from the scattering approach
\cite{Lambrecht2006,Emig2007} were limited to moderate values of $R/L$.
Results for typical experimental conditions were finally derived by developing
the scattering operator describing a round trip between plane and sphere in a
symmetrical form \cite{Hartmann2018,Hartmann2017}.

Analytical results complement the numerical work by bringing information on the
nature of PFA and its leading-order correction. In a previous paper
\cite{Spreng2018}, we have shown that the PFA result for the interaction
between two spheres is obtained by taking the geometric optical WKB Mie scattering
amplitude and using the saddle-point approximation when computing multiple
round trips between the two surfaces. The saddle point corresponds to the
condition of specular reflection at the tangent plane to the sphere at the
point of closest approach. In this paper, we show that the leading-order
correction to PFA for the plane-sphere geometry results from two independent
effects. The largest contribution arises from diffraction, which corrects the
WKB scattering amplitude taken at the saddle point. The second contribution
results from the correction to the saddle-point approximation. It accounts for
the variation of the geometric optical WKB reflection operator within the narrow
interval defined by the condition of specular reflection at the vicinity of the
point of closest approach.

We develop the scattering formula~\cite{Lambrecht2006,Emig2007} in the
plane-wave basis and expand the Casimir energy as a sum over multiple
round trips between plane and sphere, which is computed analytically for large
values of $R/L.$ As opposed to semiclassical derivations in the position
representation \cite{Schaden1998,Jaffe2004,Scardicchio2005}, the vicinity of
the PFA regime defines a narrow interval in Fourier (momentum) space, thus
allowing us to employ a saddle-point approach. Moreover, the momentum
representation provides a direct connection with the geometrical optics picture and
allows for the use of known results in semiclassical optics
\cite{Nussenzveig1992, Grandy2005}. For simplicity, we consider perfect metals
at zero temperature, but the extension to real materials would be
straightforward.

Previous derivations of the leading-order correction for the scalar
\cite{Bordag2008} and electromagnetic \cite{Teo2011} field models were based on
asymptotic approximations in the multipolar basis. Alternatively, the
derivative expansion approach also allows for the derivation of the scalar
\cite{Fosco2011} and electromagnetic \cite{BimonteEPL2012} results. Real
materials at zero temperature were also investigated within the multipolar
\cite{Teo2013} and derivative \cite{BimonteAPL2012} approaches. A non-trivial
dependence on $L/R$ was found for finite temperatures
\cite{Bordag2010LTregime,Bimonte2012,Mazzitelli2015,Bimonte2017}, depending on
the field model and material properties.

The paper is organized in the following way. In Section~\ref{sec:energy_pwb}, we
present the basic tools for developing the scattering formula in the plane-wave
basis. Section~\ref{sec:scattering_at_the_sphere} is dedicated to the Mie
reflection operator and its semiclassical expansion. The two contributions to
the leading-order correction to PFA are derived in
Section~\ref{sec:roundtrips_asymp} and concluding remarks are presented in
Section~\ref{sec:conclusions}. Some technical aspects of our calculation are
relegated to the Appendix.

\section{Casimir energy in the plane-wave basis}
\label{sec:energy_pwb}

\begin{figure}
 \begin{center}
  \includegraphics[scale=1]{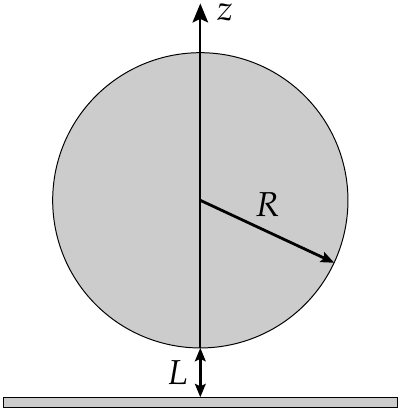}
 \end{center}
 \caption{Sphere of radius $R$ and plate separated by a distance $L$.}
 \label{fig:geometry}
\end{figure}

In this section, we develop the scattering formula for the Casimir energy in
terms of the plane-wave basis. We consider a sphere and a plate as indicated in
Fig.~\ref{fig:geometry}. We have chosen the $z$-axis as the axis of symmetry
while the plate lies in the $xy$--plane. The distance between the sphere's
center and the plate is given by $L+R$.

The plane-wave basis for the electromagnetic field is characterized by
the wave vector $\mathbf{K} = (K_x, K_y, K_z)$ and the polarization $p.$  In the region
between sphere and plate which we assume to be vacuum, the frequency of the
plane wave is determined by means of the dispersion relation $\omega =c\vert
\mathbf{K}\vert$ where $c$ is the vacuum speed of light.

As the frequency of a plane wave remains unchanged during a round trip between
sphere and plate, it is convenient to employ the
so-called angular spectral representation \cite{NietoVesperinas}. Here, the
three-dimensional wave vector $\mathbf{K}$ is replaced by the frequency $\omega$
and the two-dimensional
 projection  $\mathbf{k} = (K_x, K_y, 0)$ onto the $xy$--plane.
 Since $\omega$ and $\mathbf{k}$ only allow to determine
the modulus of $K_z$, we need to introduce the sense of propagation along the
$z$-axis, $\phi=\pm1$, so that
\begin{equation}
 K_z=\phi k_z
\end{equation}
with
\begin{equation}
 k_z = \left(\dfrac{\omega^2}{c^2}-\mathbf{k}^2\right)^{1/2}\,.
\end{equation}
A basis state within the angular spectral representation is then denoted as
$\vert \omega, \mathbf{k}, p, \phi\rangle$.

The polarization $p$ is either transverse electric (TE) or transverse magnetic
(TM) with respect to the Fresnel plane spanned by the vectors $\mathbf{K}$ and
$\mathbf{\hat z},$ with the latter defining the normal to the planar surface.
The polarization unit vectors are then defined by
\begin{equation}
 \label{eq:fresnel_basis}
 \begin{aligned}
  \hat{\boldsymbol{\epsilon}}_{\mathrm{TE}} &= \frac{\hat{\mathbf{z}}\times\hat{\mathbf{K}}}
                                               {|\hat{\mathbf{z}}\times\hat{\mathbf{K}|}}\\
  \hat{\boldsymbol{\epsilon}}_{\mathrm{TM}} &= \hat{\boldsymbol{\epsilon}}_{\mathrm{TE}}\times\hat{\mathbf{K}}
 \end{aligned}
\end{equation}
with the unit vector $\hat{\mathbf{K}} = \mathbf{K}/\vert\mathbf{K}\vert$.

In the position representation a plane wave is now given by
\begin{equation}
 \label{eq:planewave_definition}
 \langle x,y,z|\omega,\mathbf{k},p,\phi\rangle = \hat{\boldsymbol{\epsilon}}_p\left(\frac{1}{2\pi}
	\left|\frac{\omega}{ck_z}\right|\right)^{1/2} \exp\!\big[i(\mathbf{k}\cdot\mathbf{r} + \phi k_zz)\big]
\end{equation}
where $\mathbf{r} = (x,y,0)$ is the projection of the position vector onto
the $xy$--plane. The normalization prefactor is specific for the angular spectral
representation.

Within the scattering approach to the Casimir effect, it is convenient to express
the Casimir energy in terms of the imaginary frequency $\xi=-i\omega$ and to
introduce an imaginary wave vector component along the $z$-direction, $\kappa=-ik_z$.
The corresponding dispersion relation reads
\begin{equation}
 \label{eq:imag_dispersion}
 \xi^2 = c^2(\kappa^2-\mathbf{k}^2)\,.
\end{equation}
At zero temperature, the Casimir energy is then found as
an integral over imaginary frequencies~\cite{Lambrecht2006,Emig2007}:
\begin{equation}
 \label{eq:casimir_energy}
 \mathcal{E} = \hbar\int_0^\infty\frac{\mathrm{d}\xi}{2\pi}
	       \trace\log \big(1-\mathcal{M}(\xi)\big)\,.
\end{equation}
Here, $\mathcal{M}$ is an operator describing the round trip of a wave between the
objects involved. For our geometry consisting of a sphere and a plate as shown in
Fig.~\ref{fig:geometry}, we specifically have
\begin{equation}
 \label{eq:round_trip_operator}
 \mathcal{M} = \mathcal{T}_\text{PS} \mathcal{R}_\text{S}
	       \mathcal{T}_\text{SP} \mathcal{R}_\text{P}\,.
\end{equation}
$\mathcal{R}_\text{S}$ and $\mathcal{R}_\text{P}$ describe reflection at
sphere and plate, respectively. The operators $\mathcal{T}_\text{SP}$
and $\mathcal{T}_\text{PS}$ perform a translation over a distance $L+R$ along
the positive and negative $z$ direction, respectively. They are needed
to perform the transition between a reference frame situated at the sphere
center and another reference frame with the origin on top of the plate.

Within the plane-wave basis, the translation operators $\mathcal{T}_\text{PS}$
and $\mathcal{T}_\text{SP}$ are diagonal, contributing a factor
$\exp\left(-\kappa(L+R)\right)$ each. The reflection operator
$\mathcal{R}_\text{P}$ at the plane is diagonal as well with matrix elements
given by the Fresnel coefficients when taking the polarization basis defined by
(\ref{eq:fresnel_basis}). For the case of perfect reflectors considered here,
the matrix elements correspond to the reflection coefficients $r_\text{TM} = 1$
and $r_\text{TE}=-1$.

The Mie reflection operator $\mathcal{R}_\text{S}$ at the sphere requires more
attention because it couples different polarizations and values of
$\mathbf{k}$. The detailed form of the corresponding matrix elements will be
discussed in Section~\ref{sec:scattering_at_the_sphere}.

We make the Casimir energy (\ref{eq:casimir_energy}) amenable to an analytical
treatment by expanding the logarithm into a Mercator series
\begin{equation}\label{eq:energy_round_trips}
\mathcal{E} = -\hbar\sum_{r=1}^{\infty}\frac{1}{r}
	       \int_0^\infty \frac{\mathrm{d}\xi}{2\pi}\trace\mathcal{M}^r
\end{equation}
which physically implies a decomposition into terms with a specific
number $r$ of round trips between sphere and plane. In the plane-wave
representation, the trace reads
\begin{multline}\label{eq:trMr}
\trace\mathcal{M}^r = \sum_{p_0,\dots,p_{r-1}}
   \int\frac{\mathrm{d}\mathbf{k}_0\dots\mathrm{d}\mathbf{k}_{r-1}}{(2\pi)^{2r}}
                \prod_{j=0}^{r-1}e^{-2\kappa_j(L+R)}r_{p_j} \\
   \times \langle\mathbf{k}_{j+1},p_{j+1},-|\mathcal{R}_\mathrm{S}|\mathbf{k}_j,p_j,+\rangle
\end{multline}
where we sum over all intermediate polarizations $p_j$ with corresponding
reflection coefficients $r_{p_j}$ and integrate over all intermediate values of
the transversal wave vector $\mathbf{k}_j$. Here and in the following, we use a
cyclic index convention where $j=r$ is equivalent to $j=0$.

\section{Scattering at the sphere}
\label{sec:scattering_at_the_sphere}

\subsection{Exact matrix elements}
The remaining part to be specified in the decomposition of the Casimir
energy (\ref{eq:energy_round_trips}) are the matrix elements of the reflection
operator $\mathcal{R}_\mathrm{S}$ at the sphere. Because of our choice of
the plane-wave basis, this operator is the only one appearing in the round-trip
operator (\ref{eq:round_trip_operator}) leading to non-diagonal matrix elements.

An incident plane wave with wave vector $\mathbf{K}^\text{(in)}$ will be
scattered by a sphere into a superposition of plane waves with arbitrary wave
vectors $\mathbf{K}^\text{(out)}$. Here, we will consider a specific pair of
incident and scattered wave vectors $\mathbf{K}^\text{(in)}$ and
$\mathbf{K}^\text{(out)}$, respectively, which span the so-called scattering
plane. As long as the plate is not part of the scattering geometry, it is
advantageous to employ the polarization vectors with respect to the scattering
plane defined as
\begin{equation}
 \begin{aligned}
  \label{eq:scattering_basis}
  \hat{\boldsymbol{\epsilon}}_\perp &= \frac{\hat{\mathbf{K}}^\text{(out)}
	                                     \times\hat{\mathbf{K}}^\text{(in)}}
					    {|\hat{\mathbf{K}}^\text{(out)}
					     \times\hat{\mathbf{K}}^\text{(in)}|}\\
	 \hat{\boldsymbol{\epsilon}}_\parallel^\text{(in)} &=
	    \hat{\boldsymbol{\epsilon}}_\perp\times\hat{\mathbf{K}}^\text{(in)}\\
	 \hat{\boldsymbol{\epsilon}}_\parallel^\text{(out)} &=
	    \hat{\boldsymbol{\epsilon}}_\perp\times\hat{\mathbf{K}}^\text{(out)}\,.
 \end{aligned}
\end{equation}
The polarization vector $\hat{\boldsymbol{\epsilon}}_\perp$ is used for, both,
the incident and the scattered wave, while in general
$\hat{\boldsymbol{\epsilon}}_\parallel$ points in different directions in the
two scattering channels.

Since the polarization in the basis (\ref{eq:scattering_basis}) is conserved during
the scattering at a sphere, $ \mathcal{R}_\text{S}$ is block diagonal with matrix elements
\begin{equation}
 \label{eq:mie_matrixelements}
 \begin{aligned}
  \langle\mathbf{K}^\text{(out)},\perp\vert\mathcal{R}_\mathrm{S}\vert\mathbf{K}^\text{(in)},\perp\rangle &=
  \frac{2\pi c}{\xi\kappa^\text{(out)}}S_{\perp}\\
  \langle\mathbf{K}^\text{(out)},\parallel\vert\mathcal{R}_\mathrm{S}\vert\mathbf{K}^\text{(in)},\parallel\rangle &=
  \frac{2\pi c}{\xi\kappa^\text{(out)}}S_{\parallel}
 \end{aligned}
\end{equation}
where the scattering amplitudes are given by
\begin{align}
S_{\perp} &= \sum_{\ell=1}^\infty \frac{2\ell+1}{\ell(\ell+1)} \left[ a_\ell\pi_\ell(\cos\Theta)  + b_\ell\tau_\ell(\cos\Theta) \right] \label{eq:s1}\\
S_{\parallel} &= \sum_{\ell=1}^\infty \frac{2\ell+1}{\ell(\ell+1)} \left[ a_\ell\tau_\ell(\cos\Theta) + b_\ell\pi_\ell(\cos\Theta) \right]\,. \label{eq:s2}
\end{align}
Here, $a_\ell$ and $b_\ell$ are the Mie coefficients and the functions
$\pi_\ell$ and $\tau_\ell$ can be expressed in terms of Legendre polynomials
and depend on the scattering angle defined through the relation
\begin{equation}
\label{eq:def_Theta}
 \cos(\Theta) = -\frac{c^2}{\xi^2} \left( \kappa^\text{(in)}\kappa^\text{(out)}
	                  + \mathbf{k}^\text{(in)}\cdot\mathbf{k}^\text{(out)}\right) \,.
\end{equation}
The functions $a_\ell$, $b_\ell$, $\pi_\ell$, and $\tau_\ell$ are defined in
Ref.~\cite{BohrenHuffman}. For the purpose of our considerations, we will not
need the explicit expressions.

\begin{figure}
 \begin{center}
  \includegraphics[scale=0.35]{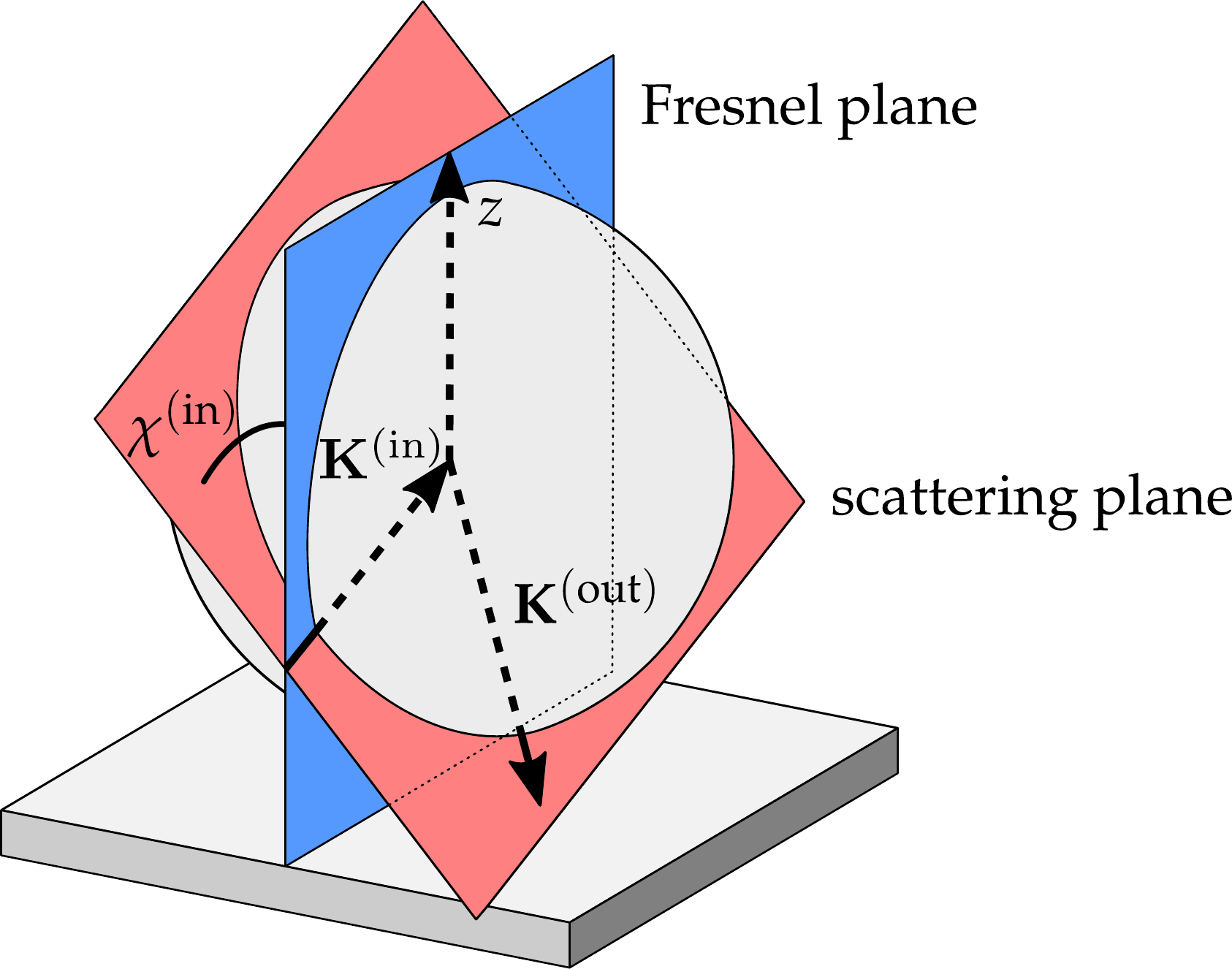}
 \end{center}
 \caption{The Fresnel plane for the incoming wave vector $\mathbf{K}^{(\mathrm{in})}$
	  in general does not coincide with the scattering plane. The two planes are at
	  an angle $\chi^{(\mathrm{in})}$. The corresponding Fresnel plane for the outgoing
	  wave vector $\mathbf{K}^{(\mathrm{out})}$ is not shown.}
 \label{fig:planes}
\end{figure}

So far, the matrix elements are expressed in the polarization basis
(\ref{eq:scattering_basis}) related to the scattering plane spanned by the
vectors $\mathbf{K}^\text{(in)}$ and $\mathbf{K}^\text{(out)}$ as displayed in
Fig.~\ref{fig:planes}. In view of our sphere-plate geometry, it is more
suitable to make use of the polarization basis (\ref{eq:fresnel_basis})
associated with the Fresnel plane spanned by the vectors
$\mathbf{K}^\text{(in)}$ and $\mathbf{\hat z}$ shown in Fig.~\ref{fig:planes},
which is tilted by an angle $\chi^\text{(in)}$ with respect to the scattering
plane. Likewise, the Fresnel plane associated to the outgoing wave vector
$\mathbf{K}^\text{(out)}$ is tilted with respect to the scattering plane by an
angle $\chi^\text{(out)}.$

The comparison between the two polarization bases (\ref{eq:fresnel_basis}) and
(\ref{eq:scattering_basis}) allows us to connect the results
(\ref{eq:mie_matrixelements}) of the Mie theory with our sphere-plate
scattering geometry. The matrix elements of  $\mathcal{R}_\text{S}$ in the Fresnel polarization basis (\ref{eq:fresnel_basis}) are given by
\begin{align}
\label{eq:matrix_element_1}
\braket{\mathbf{k}^\text{(out)},\mathrm{TM},-|\mathcal{R}_\mathrm{S}|\mathbf{k}^\text{(in)},\mathrm{TM},+}
  &=  \frac{2\pi c}{\xi \kappa^\text{(out)}} (AS_{\parallel}+BS_{\perp})  \\
\label{eq:matrix_element_2}
\braket{\mathbf{k}^\text{(out)}, \mathrm{TE}, - | \mathcal{R}_\mathrm{S} | \mathbf{k}^\text{(in)}, \mathrm{TE}, +}
  &=  \frac{2\pi c}{\xi \kappa^\text{(out)}} (AS_{\perp}+BS_{\parallel}) \\
\label{eq:matrix_element_3}
\braket{\mathbf{k}^\text{(out)}, \mathrm{TM}, - | \mathcal{R}_\mathrm{S} | \mathbf{k}^\text{(in)}, \mathrm{TE}, +}
  &= -\frac{2\pi c}{\xi \kappa^\text{(out)}} (CS_{\perp}+DS_{\parallel}) \\
\label{eq:matrix_element_4}
\braket{\mathbf{k}^\text{(out)}, \mathrm{TE}, - | \mathcal{R}_\mathrm{S} | \mathbf{k}^\text{(in)}, \mathrm{TM}, +}
  &=  \frac{2\pi c}{\xi \kappa^\text{(out)}} (CS_{\parallel}+DS_{\perp}) \,.
\end{align}
The coefficients $A$, $B$, $C$, and $D$ arise due to the fact that in general
the scattering plane and the Fresnel planes do not coincide, as illustrated in Fig.~\ref{fig:planes}. They can be cast
into the form
\begin{equation}
\label{eq:abcd}
\begin{aligned}
A &=  \cos(\chi^\text{(out)})\cos(\chi^\text{(in)}) \\
B &=  \sin(\chi^\text{(out)})\sin(\chi^\text{(in)}) \\
C &=  \sin(\chi^\text{(out)})\cos(\chi^\text{(in)}) \\
D &= -\cos(\chi^\text{(out)})\sin(\chi^\text{(in)}) \,,
\end{aligned}
\end{equation}
where the angles between the Fresnel planes and the scattering plane are defined by
the following relations
\begin{equation}
\begin{aligned}
\label{eq:angle_equations}
\cos(\chi^\text{(in)}) &= \hat{\boldsymbol{\epsilon}}_\mathrm{TE}(\mathbf{K}^\text{(in)})\cdot
	                     \hat{\boldsymbol{\epsilon}}_\mathrm{\perp}\\
\cos(\chi^\text{(out)}) &= \hat{\boldsymbol{\epsilon}}_\mathrm{TE}(\mathbf{K}^\text{(out)})\cdot
	                      \hat{\boldsymbol{\epsilon}}_\mathrm{\perp}\,.
\end{aligned}
\end{equation}

A special situation occurs when
$\mathbf{K}^{(\mathrm{out})}$ is contained in the Fresnel plane associated to $\mathbf{K}^{(\mathrm{in})}.$
In this case, the two Fresnel planes coincide with the scattering plane and
the angles $\chi^{\text{(in)}}$ and $\chi^{\text{(out)}}$
vanish. According to (\ref{eq:angle_equations}), we then find
\begin{equation}
\label{eq:ABCD_SP}
A=1, \quad B=C=D=0 \,.
\end{equation}
In this case, the matrix elements  (\ref{eq:matrix_element_1})-(\ref{eq:matrix_element_4}) become diagonal and essentially reduce to the
scattering amplitudes (\ref{eq:s1}) and (\ref{eq:s2}). It is this special
scattering situation which dominates the Casimir energy within the proximity
force approximation \cite{Spreng2018}.

\subsection{Scattering at large spheres}

In  order to derive the leading correction to the
proximity force approximation, we need the asymptotic expansions of the exact
matrix elements (\ref{eq:matrix_element_1})--(\ref{eq:matrix_element_4}) valid for large $R.$
The expansions for the  scattering amplitudes $S_{\perp}$ and $S_{\parallel}$
are well known when taking real frequencies $\omega\gg c/R$ \cite{Nussenzveig1969}.
If the close vicinity of the forward direction is excluded, they
 are obtained within the WKB approximation.

It turns out that the asymptotic imaginary-frequency expression can be obtained
by simply reexpressing the real-frequency expression in terms of imaginary
frequencies \cite{Spreng2018}. The asymptotic scattering amplitudes including
the leading correction in $1/R$ can be expressed as
\begin{equation}
 \label{eq:SA_with_corrections}
 S_p = S_p^\text{WKB}\left(1 + \frac{1}{R}s_p + \mathcal{O}\left(R^{-2}\right)\right) \,.
\end{equation}
Here, the leading WKB term is given by
\begin{equation}
\label{eq:S_WKB}
S^\text{WKB}_p = (-1)^p\frac{\xi R}{2 c} \exp\left[\frac{2\xi R}{c}\sin\left(\frac{\Theta}{2}\right)\right],
\end{equation}
where the scattering angle $\Theta$ is given by (\ref{eq:def_Theta}), and
 $p=1(2)$ stands for $\perp$ ($\parallel$) polarization.
The leading corrections in (\ref{eq:SA_with_corrections}) in order $1/R$
are found as \cite{Grandy2005}
\begin{equation}
\label{eq:correction_SA}
\begin{aligned}
s_{\perp} &= \frac{c}{2\xi} \frac{\cos(\Theta)}{\sin^3(\Theta/2)} \\
s_{\parallel} &= -\frac{c}{2\xi} \frac{1}{\sin^3(\Theta/2)} \,.
\end{aligned}
\end{equation}

It should be noted that the leading-order term (\ref{eq:S_WKB}) in real
frequencies has a clear interpretation in terms of geometrical optics \cite{Nussenzveig1992}. While the proximity force approximation of
the Casimir effect can thus be understood in terms of geometrical optics, the
leading corrections derived here require us to
take the diffraction correction  (\ref{eq:correction_SA}) into account.

Plugging the asymptotic expansion of the scattering amplitudes (\ref{eq:SA_with_corrections})
into the matrix elements (\ref{eq:matrix_element_1})--(\ref{eq:matrix_element_4}),
the asymptotic expansion of the matrix elements up to order $1/R$ can be summarized as
\begin{multline}
\label{matrix_element_asymptotic}
\braket{\mathbf{k}^\text{(out)}, p^\text{(out)},-|\mathcal{R}_\mathrm{S}|\mathbf{k}^\text{(in)},p^\text{(in)},+}\\
\simeq  \frac{\pi R}{\kappa^\text{(out)}} \exp\left[\frac{2\xi R}{c}\sin\left(\frac{\Theta}{2}\right)\right] \rho_{p^\text{(out)},p^\text{(in)}}
\end{multline}
with
\begin{equation}
\begin{aligned}
\label{eq:matrix_element_asym_ABCD}
\rho_{\mathrm{TM},\mathrm{TM}} &=  \left(A - B\right) + \frac{1}{R}\left(A s_{\parallel} - B s_{\perp} \right) \\
\rho_{\mathrm{TE},\mathrm{TE}} &= -\left(A - B\right) - \frac{1}{R}\left(A s_{\perp} - B s_{\parallel} \right) \\
\rho_{\mathrm{TE},\mathrm{TM}} &=  \left(C - D\right) + \frac{1}{R}\left(C s_{\perp} - D s_{\parallel} \right) \\
\rho_{\mathrm{TM},\mathrm{TE}} &=  \left(C - D\right) + \frac{1}{R}\left(C s_{\parallel} - D s_{\perp} \right)\,.
\end{aligned}
\end{equation}

\section{Asymptotic expansion of the Casimir energy}
\label{sec:roundtrips_asymp}

For large sphere radius, the Casimir energy can be expressed in the form
\begin{equation}
 \label{eq:def_beta1}
 \mathcal{E} = \mathcal{E}_\mathrm{PFA}\left(1+\beta_1\frac{L}{R}+o\left(R^{-1}\right)\right)\,.
\end{equation}
The well-known PFA result
\begin{equation}
 \label{eq:PFA_result}
 \mathcal{E}_\mathrm{PFA} = -\frac{\hbar c\pi^3 R}{720L^2}
\end{equation}
was shown to result from the leading order of a saddle-point evaluation of the
round-trip decomposition (\ref{eq:energy_round_trips}) \cite{Spreng2018}.
Here, we will focus on the evaluation of the constant $\beta_1$.

There exist two sources for corrections of the order $1/R$ and we therefore write
\begin{equation}
 \beta_1 = \beta_\mathrm{d}+\beta_\mathrm{go}
\end{equation}
where the indices `d' and `go' refer to diffraction and geometrical optics,
respectively.  The first contribution, $\beta_\mathrm{d}$, arises from the
diffractive correction to the scattering amplitude, i.e.\ the term involving
$s_p$ in (\ref{eq:SA_with_corrections}). The second contribution, $\beta_\mathrm{go}$,
is obtained by evaluating the saddle-point approximation to the next-to-leading order,
referred to as NTLO in the following.

The central object for obtaining an asymptotic expansion of the Casimir energy
is the trace (\ref{eq:trMr}) of the $r$-th power of the round-trip operator.
Together with the asymptotic expansion of the matrix elements of $\mathcal{R}_\mathrm{S}$
(\ref{matrix_element_asymptotic}), the trace can be brought into the form of
the $2r$-dimensional integral
\begin{equation}
\label{eq:rloops}
\trace\mathcal{M}^r \simeq \left(\frac{R}{4\pi}\right)^r\int {\mathrm{d}}\mathbf{k}_0 \dots \mathrm{d}\mathbf{k}_{r-1}\,g(\mathbf{k}_0,\dots,\mathbf{k}_{r-1}) e^{-R f(\mathbf{k}_0,\dots,\mathbf{k}_{r-1})}
\end{equation}
which is suitable for an evaluation using the saddle-point approximation.
Here, we introduced the function
\begin{equation}
 \label{eq:g_definition}
 g(\mathbf{k}_0,\dots,\mathbf{k}_{r-1}) = \sum_{p_0,\dots,p_{r-1}}\prod_{j=0}^{r-1} (-1)^{p_j}
 \frac{e^{-2\kappa_jL}}{\kappa_j}\rho_{p_{j+1},p_j} \,.
\end{equation}
The factor $(-1)^{p_j}$ represents the Fresnel coefficient $r_{p_j}$ for polarization
$p_j=1$ (TE) or $p_j=2$ (TM), which accounts for reflection at the plate. The function in the exponent in (\ref{eq:rloops}) is given by
\begin{equation}
 \label{eq:f_definition}
 f(\mathbf{k}_0,\dots,\mathbf{k}_{r-1}) = \sum_{j = 0}^{r-1} \eta_{j,j+1}
\end{equation}
where
\begin{equation}
 \label{eq:eta}
 \eta_{j,j+1} = \kappa_j + \kappa_{j+1} - \left[2\left(\frac{\xi^2}{c^2} +
	\kappa_j\kappa_{j+1}+\mathbf{k}_{j}\cdot\mathbf{k}_{j+1} \right)\right]^{1/2} \,.
\end{equation}
While the first two terms in (\ref{eq:eta}) are related to the translation from
the sphere to the plate and back, the last term is associated to the phase upon reflection
at the sphere within the WKB approximation.

\subsection{Saddle-point approximation and its leading order correction}
\label{subsec:SPA_and_beyond}

We will now derive the saddle-point approximation including the NTLO for the
integrals (\ref{eq:rloops}). In the absence of relevant boundary terms, the
dominant contribution to the integrals arises from one or more saddle points
where the gradient of the function $f$ in the exponent vanishes. The
vicinity of the saddle point is characterized by the Hessian matrix
containing second derivatives of $f$. While in leading order, only the
determinant of the Hessian matrix enters, the NTLO requires the knowledge of
the inverse of the Hessian matrix.  Therefore, before discussing the NTLO, we
first need to analyze the Hessian matrix. Luckily, it will turn out that it
allows for an analytical diagonalization
in the problem at hand.

For the function $f$ defined by (\ref{eq:f_definition}) and (\ref{eq:eta}), it
is straightforward to calculate the gradient and to determine the saddle
points. In fact, one can show that there exists a family of saddle points
\begin{equation}
\label{eq:SP_condition}
\mathbf{k}_0 = \dots = \mathbf{k}_{r-1} \equiv \mathbf{k}_\mathrm{sp}
\end{equation}
parametrized by $\mathbf{k}_\mathrm{sp}$. As a consequence, at the saddle
points the scattering plane and the Fresnel planes coincide and we have
$\chi^\text{(in)}=\chi^\text{(out)}=0$ for each reflection at the sphere.

On the saddle-point manifold, the Hessian matrix can be brought into
block-diagonal form
\begin{equation}
\mathsf{H}= \begin{pmatrix}
\mathsf{H}_{xx} & 0 \\
0               & \mathsf{H}_{yy}
\end{pmatrix}
\end{equation}
by arranging rows and columns in the order of
$(k_{0,x},\dots,k_{r-1,x},k_{0,y},\dots,k_{r-1,y})$. The matrix blocks are
given by the second derivative of $f$ evaluated at the saddle point
\begin{equation}
 \label{eq:hessian}
 \big(\mathsf{H}_{xx}\big)_{ij} = \left.\frac{\partial^2f}{\partial k_{i,x}\partial k_{j,x}}
				  \right\vert_{\mathrm{sp}}
\end{equation}
with a corresponding expression for $\mathsf{H}_{yy}$. Due to the block
structure of the Hessian matrix, we can perform the integrations over the $x$-
and $y$-components of the wave vectors separately.

The blocks of the Hessian matrix can be expressed as
$\mathsf{H}_{xx} = \mathsf{H}_{yy} = (1/2\kappa_\mathrm{sp})\Gamma_r$ with
the $r\times r$ circulant matrix
\begin{equation}
\label{eq:Gamma1}
\Gamma_r = \begin{pmatrix}
2 & -1 &&& -1\\
-1 & 2 & -1 && \\
& -1 & \ddots & \ddots &\\
& & \ddots & \ddots & -1\\
-1& & & -1 & 2 \\
\end{pmatrix}
\end{equation}
where the matrix elements not shown are zero. $\kappa_\text{sp}$ is obtained
from $\mathbf{k}_\mathrm{sp}$ by means of the dispersion relation
(\ref{eq:imag_dispersion}). In the special case of two round trips, we have
\begin{equation}
\Gamma_2 = \begin{pmatrix}
1 & -1 \\
-1 & 1
\end{pmatrix}\,,
\end{equation}
while for $r=1$ we have $f\equiv0$.

It is now convenient to introduce transformed variables $v$ through
\begin{equation}
 \label{eq:transformation}
 k_{j,x} = \sum_{l=0}^{r-1}\mathsf{W}_{jl} v_{l,x}
\end{equation}
with
\begin{equation}
\label{eq:transformation_matrix}
 \mathsf{W}_{jl}= \frac{1}{\sqrt{r}}\exp\left(\frac{2\pi i}{r}jl\right)\,.
\end{equation}
After the transformation, the blocks of the Hessian matrix are of counter-diagonal form
\begin{equation}
 \label{eq:WHW}
 \big(\mathsf{W}^T\mathsf{H}_{xx}\mathsf{W}\big)_{jl} = \lambda_j\delta_{j,r-l}
\end{equation}
with the eigenvalues
\begin{equation}
 \label{eq:eigenvalues}
 \lambda_j = \frac{2}{\kappa_\text{sp}}\sin^2\left(\frac{\pi j}{r}\right)
\end{equation}
and $j=0, 1,\ldots, r-1$. Since $\mathsf{H}_{xx} = \mathsf{H}_{yy}$, the same
procedure is applied to the $y$-components as well.

Both blocks of the Hessian matrix contain one vanishing eigenvalue associated
with the one-dimensional family of saddle points (\ref{eq:SP_condition}). As a
consequence, the variables $v_{0,x}$ and $v_{0,y}$ need to be integrated out
exactly. The remaining integrations over $v_{j,x}$ and $v_{j,y}$ with
$j=1,\ldots, r-1$ can be evaluated within the saddle-point approximation up to
NTLO as we will explain now.

If the sphere radius $R$ provides the largest length scale, the integrals
(\ref{eq:rloops}) are dominated by a region around the saddle point which
scales with $R^{-1/2}$ in all directions except for the direction of the
saddle-point family. In order to obtain the NTLO contribution, we thus need to
expand the prefactor $g$ up to second order around the saddle point while the
function $f$ appearing in the exponent needs to be expanded up to fourth order
in deviations from the saddle point.  Keeping the second-order term in the
exponent and expanding the remaining exponential up to fourth order in the
variables, we are left with integrands of Gaussian form multiplied by
polynomials. Evaluating the corresponding integrals we find
\begin{equation}
 \label{eq:asym_formula}
 \trace\mathcal{M}^r = \frac{R}{2r}\int_{\xi/c}^\infty\mathrm{d}\kappa_\mathrm{sp} \, \kappa_\mathrm{sp}^r
	\left[F_0+\frac{1}{R}F_1+o\left(R^{-1}\right)\right] \,.
\end{equation}
In deriving this result, we made use of the fact that $f\vert_\text{sp}=0$ and
that the product of the non-vanishing eigenvalues (\ref{eq:eigenvalues}) of the
Hessian yields
\begin{equation}
 \prod_{j=1}^{r-1} \frac{1}{\lambda_j} = \frac{(2\kappa_\mathrm{sp})^{r-1}}{r^2} \,.
\end{equation}
Finally, we transformed from the variables $v_{0,x}$ and $v_{0,y}$ back to
the original wave vector at the saddle point and used (\ref{eq:imag_dispersion})
to express the integral in terms of $\kappa_\mathrm{sp}$.

Up to NTLO, the integrand in (\ref{eq:asym_formula}) is specified by
\begin{equation}
 \label{eq:F0}
 F_0 = g\vert_\mathrm{sp}
\end{equation}
and
\begin{equation}
\begin{aligned}
 \label{eq:F1_intermediate}
 F_1 &= g\vert_\mathrm{sp}\left(\sum_{ijk}\frac{2f_{ijk}f_{\bar i\,\bar j\,\bar k}
	                                     +3f_{ij\bar j}f_{\bar ik\bar k}}{24\lambda_i\lambda_j\lambda_k}
			  -\sum_{ij}\frac{f_{i\bar ij\bar j}}{8\lambda_i\lambda_j}\right)\\
 &\quad+\sum_{ij}\frac{g_if_{\bar ij\bar j}}{2\lambda_i\lambda_j}
	  +\sum_i\frac{g_{i\bar i}}{2\lambda_i}\,.
\end{aligned}
\end{equation}
The eigenvalues $\lambda_i$ have been defined in (\ref{eq:eigenvalues}). For the indices,
we use the short-hand notation $\bar i = r-i$ and the summation over the indices implies
a summation also over the corresponding components $x$ and $y$. Finally, the indices at
the functions $f$ and $g$ denote derivatives with respect to the corresponding components
of $v$ evaluated at the saddle point.

A closer analysis reveals that two of the terms in (\ref{eq:F1_intermediate}) vanish because the functions
$f$ and $g$ are symmetric with respect to their arguments. Let us consider the fourth
term in (\ref{eq:F1_intermediate}) and specifically the first derivative
\begin{equation}
 \left.\frac{\partial g}{\partial v_{i,x}}\right\vert_\mathrm{sp} =
 \sum_{l=0}^{r-1}W_{il}\left.\frac{\partial g}{\partial k_{l,x}}\right\vert_\mathrm{sp}
\end{equation}
where $i\neq0$, i.e.\ we are not taking a derivative along the saddle-point manifold.
Because of the symmetry of $g$ just mentioned, the derivative on the right-hand side
evaluated at the saddle point is independent of $l$. The resulting sum over the Fourier
factors (\ref{eq:transformation_matrix}) vanishes so that $g_i=0$. Along the same lines
one can show that $f_{ij\bar j}=0$. Instead of (\ref{eq:F1_intermediate}), it is thus
sufficient to evaluate
\begin{equation}
 \label{eq:F1}
 F_1 = g\vert_\mathrm{sp}\left(\sum_{ijk}\frac{f_{ijk}f_{\bar i\,\bar j\,\bar k}}{12\lambda_i\lambda_j\lambda_k}
			  -\sum_{ij}\frac{f_{i\bar ij\bar j}}{8\lambda_i\lambda_j}\right)
              +\sum_i\frac{g_{i\bar i}}{2\lambda_i}\,.
\end{equation}

The results presented in this section take into account the corrections of order $1/R$
arising from the NTLO of the saddle-point approximation. However, it should be kept in
mind that there is a second source of corrections of order $1/R$, namely the diffractive correction in the scattering
amplitudes (\ref{eq:SA_with_corrections}). In the following subsection, we will evaluate
the leading term (\ref{eq:F0}) in the saddle-point expansion and obtain already a part of
the contributions to the NTLO Casimir term. The contributions arising from (\ref{eq:F1}) will be
discussed in the subsequent subsection.

\subsection{Leading saddle-point contribution}

At the saddle point, the projection of the wave vector onto the $xy$-plane
does not change during the scattering at the sphere and thus the scattering
plane and the Fresnel planes coincide. In particular, according to
(\ref{eq:ABCD_SP}) the coefficients $C$ and $D$ vanish and polarization mixing
does not contribute to leading order in the saddle-point approximation. As a
consequence, the leading term $F_0$ in the integrand of (\ref{eq:asym_formula})
given by (\ref{eq:F0}) can be decomposed as
\begin{equation}
\left.g\right\vert_\mathrm{sp} = g_\mathrm{TE} + g_\mathrm{TM}
\end{equation}
where in view of (\ref{eq:g_definition})
\begin{equation}\label{eq:g_p-final}
 g_p = \frac{\exp(-2\kappa_\mathrm{sp} Lr)}{\kappa_\mathrm{sp}^r}\Big(1
	        + \frac{r}{R}\left.s_p\right\vert_\mathrm{sp}\Big)\,.
\end{equation}
Expression (\ref{eq:g_p-final}) results from a sequence of $r$ reflections at the sphere with only one of them
picking the diffractive correction.
The contributions for TE and TM correspond to the terms with $p=\perp$ and
$p=\parallel$ in (\ref{eq:correction_SA}), respectively, and can be
expressed by means of (\ref{eq:def_Theta}) and (\ref{eq:imag_dispersion}) as
\begin{equation}
 \begin{aligned}
  s_\mathrm{TE}\vert_\mathrm{sp} &= \frac{1}{\kappa_\mathrm{sp}^3}\left(\frac{\xi^2}{2c^2}-
	                                             \kappa_\mathrm{sp}^2\right)\\
  s_\mathrm{TM}\vert_\mathrm{sp} &= -\frac{1}{\kappa_\mathrm{sp}^3}\frac{\xi^2}{2c^2}\,.
 \end{aligned}
\end{equation}

Carrying out the integral over $\kappa_\mathrm{sp}$ in (\ref{eq:asym_formula}),
we find for the contributions of the two polarizations arising from $F_0$
\begin{equation}
\Big(\trace\mathcal{M}^r_\mathrm{TE}\Big)_{0} = \frac{R}{L}\frac{e^{-u}}{4r^2} + \frac{1}{8} \left[\left(u^2-4\right) \text{E}_1(u)-(u-1)e^{-u}\right]
\end{equation}
and
\begin{equation}
\Big(\trace\mathcal{M}^r_\mathrm{TM}\Big)_{0} = \frac{R}{L}\frac{e^{-u}}{4r^2} -\frac{1}{8} \left[u^2 \text{E}_1(u)-(u-1) e^{-u}\right]
\end{equation}
Here, $u=2\xi{L}r/c$ and $\mathrm{E}_1$ denotes the exponential integral function~\cite{DLMF6.2}.

Evaluating the expression for the Casimir energy (\ref{eq:energy_round_trips}), we obtain
\begin{equation}
 \label{eq:e_p0}
 \mathcal{E}_{p,0} = \mathcal{E}_\mathrm{PFA}\left(\frac{1}{2}+\beta_{\mathrm{d},p}\frac{L}{R}\right)
\end{equation}
with
\begin{align}
 \beta_{\mathrm{d, TE}} &= -\frac{25}{2\pi^2}\label{eq:beta_d_te}\\
 \beta_{\mathrm{d, TM}} &= -\frac{5}{2\pi^2}\label{eq:beta_d_tm}\,.
\end{align}
The complete diffractive correction of order $1/R$ is thus quantified by
\begin{equation}
 \label{eq:beta_d}
 \beta_\mathrm{d} = -\frac{15}{\pi^2}\,.
\end{equation}

\subsection{Geometric optical correction to PFA}
\label{subsec:go_correction}

In this section, we calculate the remaining part of the correction to PFA, i.e.\
the coefficient $\beta_\mathrm{go}$. This contribution is due to the first
correction of the saddle-point approximation with the integrand $F_1$ specified
in (\ref{eq:F1}). Since these terms are already of order $1/R$, we only need to
take into account the leading-order term in the matrix elements
(\ref{eq:matrix_element_asym_ABCD}) of the reflection operator. In other words, the subleading
term in the matrix elements associated to diffraction does not contribute to order $1/R$ when computing
the integral of $F_1$ in (\ref{eq:asym_formula}).
Therefore, the term discussed in this
section can be interpreted in terms of geometrical optics.

Introducing the angle $\chi = \chi^\text{(in)} + \chi^\text{(out)}$, we obtain from
(\ref{eq:abcd}) and (\ref{eq:matrix_element_asym_ABCD}) to leading order
\begin{equation}
 \begin{aligned}
  \rho_{\mathrm{TM},\mathrm{TM}} = -\rho_{\mathrm{TE},\mathrm{TE}} &= \cos(\chi)\\
  \rho_{\mathrm{TE},\mathrm{TM}} = \rho_{\mathrm{TM},\mathrm{TE}} &= \sin(\chi)\,.
 \end{aligned}
\end{equation}
Since at the saddle points, scattering plane and Fresnel planes coincide, we
have $\chi\vert_\text{sp}=0$ as well as
$\rho_{\mathrm{TE},\mathrm{TM}}\vert_\text{sp}=
\rho_{\mathrm{TM},\mathrm{TE}}\vert_\text{sp}=0$. In view of
(\ref{eq:g_definition}), a non-vanishing tilt between scattering plane and
Fresnel planes can thus only enter through the last term in (\ref{eq:F1}).
However, this is not the case as we will show now.

In view of the trace in (\ref{eq:rloops}), the sequence of $r$ scattering processes
necessarily involves an even number of polarization changes. Since the derivatives
in $g_{i\bar i}$ appearing in (\ref{eq:F1}) have to be evaluated at the saddle point,
the relevant terms either contain zero or two polarization-mixing matrix elements.
The relevant polarization-dependent contribution in (\ref{eq:g_definition}) thus reads
\begin{multline}
\label{eq:polarizationterms}
\sum_{p_0,\ldots,p_{r-1}}\prod_{j=0}^{r-1}(-1)^{p_j}\rho_{p_{j-1}, p_j}\\
	= \prod_{i=0}^{r-1}\cos(\chi_{i+1,i})\Bigg(1 - \sum_{j>l=0}^{r-1}\tan(\chi_{j+1,j})\tan(\chi_{l+1,l})\Bigg)
\end{multline}
where $\chi_{j+1,j}$ describes the sum of the angles $\chi^\text{(in)}$ and
$\chi^\text{(out)}$ for the $j$-th scattering process.

As a single derivative of (\ref{eq:polarizationterms}) vanishes when evaluated
at the saddle point, there are two contributions to $g_{i\bar i}$. In the first
contribution, no derivative of (\ref{eq:polarizationterms}) is taken. This term can
then simply be replaced by a factor of 1 and does not account for a tilt
between the scattering plane and the Fresnel planes. The second contribution
potentially accounts for such a tilt and takes the form
\begin{equation}
 \label{eq:polarization_derivative}
 \begin{aligned}
&\left.\left(\frac{\partial^2}{\partial v_{i,x} \partial v_{r-i,x}}
\sum_{p_0,\ldots,p_{r-1}}\prod_{j=0}^{r-1}(-1)^{p_j}\rho_{p_{j-1}, p_j}
\right)\right\vert_\mathrm{sp}\\
	 &=  - \sum_j \left.\left(\frac{\partial \chi_{j+1,j}}{\partial v_{i,x}}\frac{\partial \chi_{j+1,j}}{\partial v_{r-i,x}}\right)\right\vert_\mathrm{sp}\\
	 &\qquad- \sum_{j>l} \left.\left(\frac{\partial \chi_{j+1,j}}{\partial v_{i,x}}\frac{\partial \chi_{l+1,l}}{\partial v_{r-i,x}} +\frac{\partial \chi_{j+1,j}}{\partial v_{r-i,x}}\frac{\partial \chi_{l+1,l}}{\partial v_{i,x}}\right)\right\vert_\mathrm{sp}
 \end{aligned}
\end{equation}
where the first term arises from the second derivative of one of the cosine factors
in (\ref{eq:polarizationterms}), while the second term is obtained from single
derivatives of two tangent factors. Combining the two sums allows us to express
(\ref{eq:polarization_derivative}) as a product of two first derivatives of
$\sum_{j=0}^{r-1}\chi_{j,j+1}$. As the latter quantity is symmetric in its arguments,
its first derivative vanishes at the saddle point as demonstrated above.
We conclude that the correction (\ref{eq:F1}) arises only from scattering processes
where the scattering plane is identical with the two Fresnel planes. The evaluation
of (\ref{eq:F1}) can thus be done as if there were no polarization mixing in the
Fresnel polarization basis (\ref{eq:fresnel_basis}).

For the technical details of the remaining evaluation of (\ref{eq:F1}), we
refer the reader to the appendix. After integration over the saddle-point
manifold, we find for the correction arising from $F_1$
\begin{equation}
\Big(\trace\mathcal{M}^r_p\Big)_{1} = -\frac{(r^2-1)\exp(-2Lr\xi/c)}{12r^2}
\end{equation}
which is independent of the polarization $p=\mathrm{TE},\mathrm{TM}$.
Carrying out the integration over imaginary frequencies $\xi$ in (\ref{eq:energy_round_trips}),
we obtain for the geometric optical correction
\begin{equation}
 \label{eq:beta_go}
 \beta_\text{go} = \frac{1}{3} - \frac{5}{\pi^2}
\end{equation}
to which the two polarizations contribute equally.

\begin{figure}
 \begin{center}
  \includegraphics[scale=0.3]{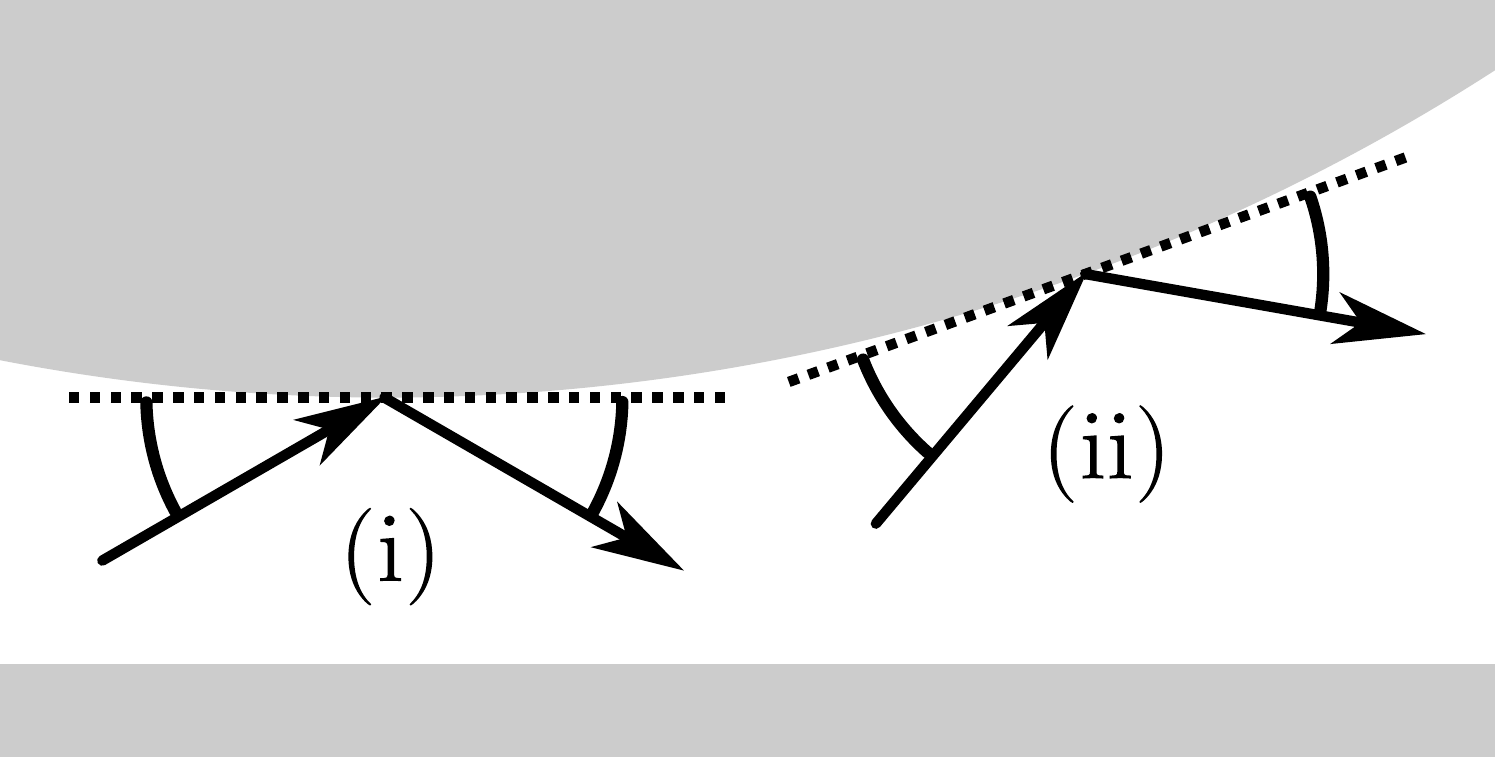}
 \end{center}
 \caption{Specular reflection at (i) a tangent plane at the bottom of the sphere and (ii)
	  at a slightly tilted tangent plane.}
 \label{fig:tilt}
\end{figure}

As shown in~\cite{Spreng2018}, the PFA result corresponds to the leading term
of the saddle-point approximation which arises from ray-optical specular
reflection at the point of the sphere closest to the plate, i.e.\ scattering
channel (i) in Fig.~\ref{fig:tilt}. The correction (\ref{eq:beta_go}) can still
be understood within geometrical optics but now the specular reflections may
also occur at tangent planes slightly tilted with respect to the plate as
illustrated by channel (ii) in Fig.~\ref{fig:tilt}.

\subsection{Total leading order correction to PFA}

The two contributions (\ref{eq:beta_d}) and (\ref{eq:beta_go}) are both negative and
add up to the NTLO correction
\begin{equation}
 \label{eq:beta1}
 \beta_1 = \frac{1}{3}-\frac{20}{\pi^2} \approx -1.693\,.
\end{equation}
This result is known from the literature \cite{Teo2011,BimonteEPL2012} but now it has
obtained a physical interpretation in terms of the relevant physical scattering processes.
Furthermore, we have found that the contribution dominating by far is due to diffraction.
According to table~\ref{table:percentage}, diffraction contributes
almost 90\% to the reduction of the Casimir energy while specular reflection
at a tangent plane inclined with respect to the plate contributes little over 10\%.
It is known that within the proximity force approximation, only a small effective
area of radius $(RL)^{1/2}$ around the point of closest approach contributes to the Casimir energy (see for instance \cite{Spreng2018}).
The rather small
contribution from geometrical optics corrections implies that the effective area
argument is barely changed by the leading corrections. Curvature effects of the
sphere manifest themselves mostly through diffraction.

\begin{table}
\centering
\begin{tabular}{| l | r | r |}
	\hline
		   & \multicolumn{1}{c|}{TE} & \multicolumn{1}{c|}{TM} \\\hline
		diffraction        & 74.8\% & 15.0\%\\
		geometrical optics &  5.1\% & 5.1\%\\
	\hline
\end{tabular}
\caption{Relative contribution of the terms arising from diffraction
	and geometrical optics for the transverse electric (TE) and
	transverse magnetic (TM) polarization to the total correction of
	order $1/R$.}
\label{table:percentage}
\end{table}

The correction $\beta_1$ can also be split according to polarization. Keeping in
mind that both polarizations contribute equally to $\beta_\text{go}$, we find
from (\ref{eq:beta_d_te}), (\ref{eq:beta_d_tm}), and (\ref{eq:beta_go})
the contributions from the transverse electric modes
\begin{equation}
 \beta_\text{TE} = \frac{1}{6}-\frac{15}{\pi^2} \approx -1.353
\end{equation}
and the transverse magnetic modes
\begin{equation}
 \beta_\text{TM} = \frac{1}{6}-\frac{5}{\pi^2} \approx -0.339\,.
\end{equation}
The TE contribution is thus about four times as large as the TM contribution.

In the literature, it has been noticed that the correction to the PFA coincides
with the sum of two scalar field contributions, namely, Dirichlet-Dirichlet
(DD) and Neumann-Neumann (NN) boundary condition \cite{Teo2011,BimonteEPL2012}.
In this case, the individual contributions are given by
\begin{equation}
 \beta_\text{DD} = \frac{1}{6}
\end{equation}
and
\begin{equation}
 \beta_\text{NN} = \frac{1}{6}-\frac{20}{\pi^2}\,.
\end{equation}
Interestingly, this decomposition is not related to the physical mechanisms
revealed in the present work.

\section{Conclusions}
\label{sec:conclusions}

We have derived the leading order correction to PFA in the plane-sphere
geometry by developing the scattering formula in the plane wave basis. The
momentum representation allows us to make a direct connection with geometrical
optics and known results in semiclassical Mie scattering. Diffraction accounts
for most of the total correction, with the TE polarization yielding a larger
contribution than the TM one. The diffraction contribution is calculated to
leading order in the saddle-point approximation, and it amounts to correcting
the condition of geometric optical specular reflection at the tangent plane at
the bottom of the sphere, i.e.\ scattering channel (i) in Fig.~\ref{fig:tilt},
by the leading order curvature effect.

The remaining part of the correction to PFA arises from the NTLO term in the
saddle-point expansion, with the round-trip operator computed within the
leading order WKB approximation. Such round trips correspond to a sequence of
geometric optical specular reflections between plane and sphere, with the
reflections at the latter taken at tangent planes which are slightly tilted
with respect to the tangent plane at the bottom of the sphere, as illustrated
by scattering channel (ii) in Fig.~\ref{fig:tilt}. Some of the scattering
channels associated to a tilted tangent plane allow for a mixing between TE and
TM polarizations, provided that the scattering plane is also tilted with
respect to the Fresnel plane as in the case shown in Fig.~\ref{fig:planes}.

It is important to understand how polarization mixing channels contribute to
the geometric optical correction to PFA since they are known to lead to
negative Casimir entropies of geometrical origin
\cite{Canaguier-Durand2010,Canaguier2010PRA,Zandi2010,Milton2015,Ingold2015,Umrath2015}.
Although the polarization mixing matrix elements provide a non-vanishing
contribution, the total correction associated to the tilt between the
scattering and Fresnel planes turns out to vanish to NTLO. In other words, the
final result for the leading order correction to PFA would be the same if the
complications associated to the difference between the Fresnel and scattering
polarization bases had been discarded from the beginning. Such remarks suggest
that an alternative derivation, in which the polarization mixing effect would
be entirely absent, would more directly lead to the leading order correction to
PFA.

\textbf{Funding.}
German Academic Exchange Service
(DAAD);
 Coordination for the Improvement of Higher
Education Personnel (CAPES-Brazil); Binational CAPES/DAAD PROBRAL collaboration program;
National Council for Scientific and Technological Development
(CNPq); National Institute of Science and Technology Complex Fluids
(INCT-FCx); Research Foundations of the States of Minas Gerais
(FAPEMIG), Rio de Janeiro (FAPERJ) and S\~ao Paulo (FAPESP) (2014/50983-3).

\appendix

\section{Derivation of the next-to-leading-order term in the saddle-point expansion}
\label{F1-calculation}

In this appendix, we present more details about the calculation of the NTLO term
in the saddle-point approximation. More specifically, we calculate the
contribution of the term
\begin{equation}
F_1 = g\vert_\text{sp}\left(\frac{D_1}{12} -\frac{D_2}{8}\right) +\frac{D_3}{2}
\end{equation}
with
\begin{equation}
\begin{aligned}
D_1 &= \sum_{\alpha, \beta, \gamma \in \{x, y\}} \sum_{i,j,l=1}^{r-1}
       \frac{1}{\lambda_i \lambda_j \lambda_l}\frac{\partial^3 f}{\partial v_{i, \alpha} \partial v_{j, \beta}
	        \partial v_{l, \gamma}}\frac{\partial^3 f}{\partial v_{\bar i, \alpha}
		\partial v_{\bar j, \beta} \partial v_{\bar l, \gamma}} \\
D_2 &= \sum_{\alpha, \beta \in \{x, y\}} \sum_{i,j=1}^{r-1} \frac{1}{\lambda_i \lambda_j}
       \frac{\partial^4 f}{\partial v_{i, \alpha} \partial v_{\bar i, \alpha} \partial v_{j, \beta}
	                   \partial v_{\bar j, \beta}} \\
D_3 &= \sum_{\alpha \in \{x, y\}} \sum_{i=1}^{r-1} \frac{1}{\lambda_i}\frac{\partial^2 g}
	    {\partial v_{i, \alpha} \partial v_{\bar i, \alpha}}\,,
\end{aligned}
\end{equation}
where we made use of the notation $\bar i = r-i$ introduced in
Section~\ref{sec:roundtrips_asymp}\ref{subsec:SPA_and_beyond}.

In the following, we demonstrate the calculation of the most complex term $D_1$.
The other terms can be computed analogously. After employing the chain rule, $D_1$ can be written as
\begin{multline}
\label{eq:D1}
D_1 = \sum_{p,q=0}^{r-1} \sum_{m,n,s=p}^{p+1} \sum_{t,u,w=q}^{q+1} a(m-t)a(n-u)a(s-w) \\ \times d_{pq}(m,n,s;t,u,w)
\end{multline}
where
\begin{equation}
a(s) = \frac{1}{r}\sum_{j=1}^{r-1} \frac{e^{2\pi i j s/r}}{\lambda_j}
\end{equation}
and
\begin{equation}\label{eq:d_pq}
d_{pq}(m,n,s;t,u,w) = \sum_{\alpha,\beta,\gamma \in \{x, y\}} \frac{\partial^3 \eta_{p,p+1}}{\partial k_{m, \alpha} \partial k_{n, \beta} \partial k_{s, \gamma}}\frac{\partial^3 \eta_{q,q+1}}{\partial k_{t, \alpha} \partial k_{u, \beta} \partial k_{w, \gamma}}\,.
\end{equation}
Using the identity \cite{Berndt2002}
\begin{equation}
\sum_{j=1}^{r-1} \frac{e^{2\pi i j s/r}}{\sin^2(\pi j /r)} = \frac{1}{3}\left(r^2 - 6 \vert s \vert r + 6 s^2 -1\right)\,,
\end{equation}
the function $a$ evaluates to
\begin{equation}
a(s) = \frac{\kappa_\mathrm{sp}}{6 r}\left(r^2 - 6sr + 6 s^2 -1\right)\,.
\end{equation}
In view of its definition through a Fourier series, this function should be understood
as $r$-periodic with $0\leq s\leq r$.

The sum over the indices $m$, $n$ and $s$ in (\ref{eq:D1}) runs only over $p$
and $p+1$ since all other partial derivatives in (\ref{eq:d_pq}) vanish. For the
same reason the indices $t$, $u$ and $w$ take only the values $q$ and $q+1$.
Thus, we sum over 64 different arguments of the function $d_{pq}$. However, there
are only three classes of arguments for which $d_{pq}$ yields a non-zero value.
These are given by
\begin{equation}
\label{eq:d_pq_d}
\begin{aligned}
d_{pq}(p,p,p;q,q,q) &= d \\
d_{pq}(p+1,p,p;q,q,q) = d_{pq}(p,p,p;q+1,q,q)&= -\frac{d}{3} \\
d_{pq}(p+1,p,p;q+1,q,q) &= \frac{d}{3}
\end{aligned}
\end{equation}
with
\begin{equation}
 d = \frac{3}{4}\frac{\mathbf{k}_\mathrm{sp}^2}{\kappa_\mathrm{sp}^6}\,.
\end{equation}
On the other hand
\begin{equation}
d_{pq}(p+1,p,p;q,q+1,q) = d_{pq}(p+1,p,p;q,q,q+1) = 0\,.
\end{equation}
All other sets of arguments can be reduced to the forms given by means of the
following rules. In each triple of arguments one can perform the replacement
$p\leftrightarrow p+1$ and/or $q\leftrightarrow q+1$ because the derivatives
are evaluated at a saddle point. In this way, at most one argument is $p+1$ or
$q+1$. Furthermore, because of commutativity of the partial derivatives, one
can permute the two triples of arguments. However, this permutation has to
be done in the same way on both triples as the derivatives are coupled through
the indices $\alpha,\beta,$ and $\gamma$ in (\ref{eq:d_pq}). In this way, the
argument $p+1$, if it exists, can be brought to the first position and we end
up with one of the sets of arguments given above.

Taking these rules into account, (\ref{eq:D1}) can be expressed as
\begin{equation}\label{eq:D1-pq-sum}
D_1 = \sum_{p,q=0}^{r-1} A(p-q)
\end{equation}
with
\begin{multline}
A(s) = d\Big(6a^3(s) + \left[a(s-1)+a(s+1)\right] \\
\times\left[a(s-1)a(s+1)-4a^2(s)\right]\Big) \,.
\end{multline}
The resulting sum over polynomials can be evaluated and one finds
\begin{equation}
D_1 = \frac{(r-2)(r-1)^2 (c^2\kappa_\mathrm{sp}^2-\xi^2)}{rc^2\kappa_\mathrm{sp}^3}\,.
\end{equation}

The calculation for $D_2$ and $D_3$ is analogous but simpler. Note, however, that
in contrast to (\ref{eq:d_pq_d}) the various types of contributions do not necessarily
differ simply by a numerical factor. Carrying out the calculation, one finds
\begin{equation}
D_2 = \frac{2(r-1)^2\left((r-2)c^2\kappa_\mathrm{sp}^2 - 3r\xi^2\right)}{3 rc^2\kappa_\mathrm{sp}^3}
\end{equation}
and
\begin{equation}
D_3 = -\frac{(r^2-1)\left(\xi^2 + L \kappa_\mathrm{sp}(c^2\kappa_\mathrm{sp}^2+\xi^2)\right)}{3c^2\kappa_\mathrm{sp}^3}
	g\vert_\mathrm{sp}\,.
\end{equation}
We recall that according to our discussion in Sec.~\ref{sec:roundtrips_asymp}\ref{subsec:go_correction}
the evaluation of $D_3$ has to be done for $\chi=0$.

The total NTLO term in the saddle-point expansion thus is given by
\begin{equation}
F_1 = - \frac{(r^2-1)\left(rL\kappa_\mathrm{sp}(c^2\kappa_\mathrm{sp}^2+\xi^2)+ \xi^2\right)}{6rc^2\kappa_\mathrm{sp}^3} g\vert_\text{sp}\,.
\end{equation}

\end{document}